\documentclass[a4,12pt]{article}
\usepackage{graphicx}
\newcommand{\be}{\begin{eqnarray}}
\newcommand{\ee}{\end{eqnarray}}
\newcommand{\NP}[1]{{\it Nucl.\ Phys.\ }{\bf #1}}
\newcommand{\PL}[1]{{\it Phys.\ Lett.\ }{\bf #1}}

\newcommand{\PR}[1]{{\it Phys.\ Rev.\ }{\bf #1}}

\newcommand{\HPA}[1]{{\it Helv.\ Phys.\ Acta.\ }{\bf #1}}
\title{A New Approach for Analytic Amplitude Calculations\\{~}}
\author{Cong-Feng Qiao
\\\\
{\small Department of Physics, Graduate School of Chinese 
Academy of Sciences,}\\
{\small YuQuan Road 19A, 100039 Beijing, China}\\
and\\
{\small The Abdus Salam International Centre for Theoretical 
Physics(ICTP),}\\
{\small Strada Costiera, 11-34014 Trieste, Italy}}
\date{}
\begin{document}
\maketitle
\thispagestyle{empty}
\begin{abstract}

We present a method for symbolic calculation of Feynman amplitudes 
for processes involving both massless and massive fermions. With this
approach fermion strings in a specific amplitude can be easily
evaluated and expressed as basic Lorentz scalars. The new approach 
renders the symbolic calculation of some complicated physical 
processes more feasible and easier, especially with the assistance 
of algebra manipulating code for computer.
 
\vspace*{7mm}

\noindent
PACS number(s) : 13.85.Qk, 14.70.Dj, 14.70.Hp
\end{abstract}

\clearpage

\section{Introduction}
In high energy collisions, generally, multi-particle final states 
are produced. To calculate a practical physical process with several
fermions in the final state, conventionally, is a tedious work to
implement, even at the tree level. One needs to square the Feynman
amplitude, sum over fermion polarizations, and to take trace for 
each possible fermion string loops. The non-Abelian nature of the 
Standard Model makes the number of independent Feynman diagrams grow 
very rapidly as the external particles increasing. Fortunately, 
due to the recent development in computer technology, a lot of time 
consuming and tedious Dirac algebra and symbolic simplification work 
can be carried out on a ordinary personal computer. Nevertheless, 
even with the up-to-date algebra manipulating programs for computer, 
many a multi-particle process tend to be still cumbersome to calculate. 
On the other hand, another drawback of conventional amplitude squaring
technique is that it loses the information on spin correlations, which
is more and more attainable in present experimental measurements. 

An alternative technique, the so-called helicity method, has been
developed pretty applicable in practice in recent years. With this
method, the independent fermion strings are contracted and reduced
to analytic expressions of four-vector products, such that the
matrix-element squaring becomes a trivial work. The development of 
helicity method has experienced a bit long history. Some early efforts 
on this can be seen in Refs. \cite{r1,r2,r3}. In recent development, 
a bunch of approaches for calculating the helicity amplitudes are
proposed in Refs. [4-14], and it should be noted the references 
given here are far from complete on this subject. 

In practice, most of physical processes involve fermion production
and decays, which give a number of independent fermion lines within 
a Feynman amplitude. The generic form of a fermion line involving a 
pair of fermions can be expressed as
\be
\label{e1}
\bar{u}(p_1,\lambda_1)\not\!{G_1}\not\!{G_2}\cdots\not\!{G_n} 
{u}(p_2,\lambda_2)\; ,
\ee
where $(p_1,\lambda_1)$ and $(p_2,\lambda_2)$ stand for the momenta
and polarizations of the external fermions; and in between of the two
spinors there is a string of Dirac matrices. Since real physical
processes are described by the matrix element square, in conventional 
method of calculating an interaction process, one squares the
amplitude, which contains more or less (\ref{e1})-like fermion lines
when there are external fermions involved, and sums over the fermion
polarizations to get fermion loop(s). Using the nature of spinor
projection operators, one then is confronted with couple of traces.
As aforementioned, with the growth of the number of external fermions
one need to evaluate a vast number of fermion loop traces. Even with
the help of computer algebra manipulation, quite often one finds
it is still a cumbersome work to perform the calculation. 

Helicity method enables one to evaluate physical amplitude directly,
resulting in a limited number of four-vector scalars. In literatures, 
there are several kinds of approaches in evaluating the fermion 
strings analytically. One of them is carried out by choosing a specific 
representation for the Dirac matrices \cite{hz}, and then get the 
explicit forms of all the related matrices. Another kind of popular 
helicity method \cite{ks,bama} is performed by introducing
extra lightlike auxiliary momenta and expressing the massive particle
spinors by the corresponding massless ones. 

In this work, we use the maximum credits of Refs. \cite{ks,bama} to
generalize a straightforward method for the evaluation of fermion 
strings. We believe this approach, which leads to a covariant form 
for the helicity amplitude, is a relatively simple and fast way in 
practical computations.

\section{Formalism}

In the conventional method one computes unpolarized cross sections 
by squaring the eq. (\ref{e1})-like amplitude while summing over 
fermion polarizations, and obtains
\be
\label{e2}
{\rm tr}[(\not\!{p_1} \pm m_1) \not\!{G_1}\not\!{G_2}
\cdots\not\!{G_n} 
(\not\!{p_2} \pm m_2) \not\!{G_n}\cdots\not\!{G_2}\not\!{G_1}]
\; ,
\ee
where $m_1$ and $m_2$ are the masses of $p_1$ and $p_2$, and the 
plus and minus signs before them correspond to particle and 
anti-particle, respectively. Taking the above trace in principle 
is not a tedious work, especially with the help of computer. However,
with the number of external fermions increasing, in conventional 
amplitude modulus square approach one encounters a prohibitive
number of eq.(\ref{e2})-like traces to evaluate, which makes the
physical computations very time consuming, even for the machine.

Noticing that the fermion string (\ref{e1}) can be rewritten as
a trace form, like
\be
\label{e3}
{\rm tr}[\not\!{G_1}\not\!{G_2}\cdots\not\!{G_n} 
{u}(p_2,\lambda_2)\otimes\bar{u}(p_1,\lambda_1)]\; ,
\ee
therefore, now, the question of how to evaluate the helicity 
amplitude becomes how to re-express 
${u}(p_2,\lambda_2)\otimes\bar{u}(p_1,\lambda_1)$ 
by basic Dirac matrices. In Ref.\cite{vewu}, Vega and Wudka 
obtained one kind of expression for it by making use of the 
Bouchiat-Michel identity \cite{bomi}. While in principle their 
result is applicable for practical calculation, we still feel 
it not so convenient for programming, since the main result of 
(9) in Ref.\cite{vewu} is not expressed in a convenient and 
covariant way for different helicities. 

The helicity method we will present is a straightforward 
generalization of the results of Refs.\cite{ks,bama}. By choosing 
two auxiliary four-vectors, $k_0$ and $k_1$, one can fix up the 
basic spinors $u(k_0,\lambda)$ by the equations:
\be
\label{e4}
{u}(k_0,\lambda)\bar{u}(k_0,\lambda) = 
\frac{1 + \lambda \gamma_5}{2}\not\!{k_0}
\ee
and
\be
\label{e5}
{u}(k_0,\lambda) = \lambda \not\!{k_1} {u}(k_0, -\lambda)\; .
\ee
Here,
\be
\label{e6}
k_0\cdot k_0 = 0,\; k_1\cdot k_1 = -1,\; k_0\cdot k_1 = 0\; . 
\ee
With the above defined basic spinors, the wavefunction of a massive 
fermion(particle or anti-particle) can be expressed as
\be
\label{e7}
{u}(p,\lambda) = 
\frac{\not\!{p} \pm m}{\sqrt{2 p\cdot k_0}}\; u(k_0, -\lambda)\; .
\ee
It is easy to check that these spinors, and their conjugations 
defined by the normal way, satisfy Dirac equations; they are also 
the eigenstates of $\gamma_5 \not\!n$ with eigenvalues of $\pm 1$. 
Here, n is the fermion polarization vector. Using Eqs.(\ref{e4}), 
(\ref{e5}), and (\ref{e7}), one can readily get the desired spinor 
products in (\ref{e3}), like
\be
\label{e8}
&&{u}(p_2,\lambda_2)\otimes \bar{u}(p_1, \lambda_1) =
\frac{\not\!{p_2} \pm m_2}{\sqrt{2 p_2\cdot k_0}}\; u(k_0, -\lambda_2)
\bar{u}(k_0, -\lambda_1) \frac{\not\!{p_1} \pm m_1}{\sqrt{2 p_1\cdot k_0}}
\nonumber \\
&&= 
\frac{\not\!{p_2} \pm m_2}{2 \sqrt{2 p_2\cdot k_0}} 
\{ (1 + \lambda_1 \lambda_2) - (\lambda_1 + \lambda_2)\gamma_5 
\nonumber \\
&&+ \not\!{k_1}[ (\lambda_1 - \lambda_2) - 
(1 - \lambda_1 \lambda_2)\gamma_5 ] \} \not\!{k_0}
\frac{\not\!{p_1} \pm m_1}{2 \sqrt{2 p_1\cdot k_0}}
\; .
\ee
In fact (\ref{e8}) was obtained in \cite{bama}, but it 
was not realized that (\ref{e8}) can be directly applied to the 
calculation of helicity amplitude, rather, there it was used for 
evaluating the segments of a splitted fermion string. 
This finding makes the computation of practical processes 
much simplified from the method of Ref. \cite{bama}.

\section{Applications and Concluding Remarks}

In arriving at eq.(\ref{e8}) we have taken no special provisos on 
the fermion mass, hence, it will remain valid for massless case. From 
(\ref{e8}) we can easily obtain the known identities of the inner 
product of two massless spinors \cite{peskin}, like 
\be
\label{e9}
\bar{u}(p, -) {u}(p', +) = [\bar{u}(p', +) {u}(p, -)]^*\; ,
\ee
\be
\label{e9p}
\bar{u}(p, +) {u}(p', -) = - \bar{u}(p', +) {u}(p, -)\; ,
\ee
and
\be
\label{e10}
|\bar{u}(p, +) {u}(p', -)|^2 = 2\; p\cdot p'\; .
\ee

When taking a specific choice for auxiliary vector $k_0$, that is
\be
\label{e11}
k_0^0 = \frac{\alpha}{m^2} (p^0 - |{\bf p}|)\; , \; {\bf k}_0 = 
\frac{\alpha}{m^2} (|\bf{p}| - p^0) \hat{\bf p}\; ,
\ee
where $\alpha$ is an arbitrary parameter and $\hat{\bf p}$ the 
unit vector of space components of momentum p, the polarization 
vector now becomes the conventional helicity vector:
\be
\label{e12}
n = (\; \frac{\bf p}{m}\; ,\; \frac{p_0}{m}\hat{\bf p}\; )\; .
\ee
Then, from (\ref{e8}) one easily finds that at helicity basis, 
the spinors $u(p,\lambda)$ satisfies the usual projection relation
\be
\label{e13}
u(p, \lambda) \bar{u}(p, \lambda) = \frac{(1 + \lambda \gamma_5 
\not\!n)}{2} (\not\!p \pm m)\; ,
\ee

As the last example, we compute the helicity amplitude for fermion 
line with a single vector current insertion. That is
\be
\label{e15}
A_\mu(p,\lambda; \; p',\lambda') = \bar{u}(p, \lambda)
\gamma_\mu {u}(p', \lambda')\; .
\ee
This kind of amplitude appears quite often in fermion production and 
decay processes, e.g., $e^+\; e^- \rightarrow f \bar{f}$. In the 
the center-of-mass(CM) system, with choosing
\be
\label{e16}
p = (p_0,0,0, p_3)\; ,\; p' = (p_0, 0,0, -p_3)\; ,\;
k_0 = (1, 0, 0, -1)\; ,\; k_1 =(0, 1, 0, 0)\; 
\ee
and using (\ref{e8}) the helicity amplitude $A_\mu$ can be can be 
evaluated as
\be
\label{e17}
A_\mu (p,\lambda; \; p',\lambda') 
&=& \frac{1}{2 m}\left\{ \frac{(1 + \lambda \lambda')}{2} 
\right. (s\; \delta_{\mu 0} - (s - 4 m^2)\; 
\delta_{\mu 3} - s k_{0\mu}) \nonumber\\
&+& (\lambda' - \lambda) m \sqrt{s - 4 m^2}\; \delta_{\mu 1}
\left. - {\rm I} (1 - \lambda \lambda') m \sqrt{s}\; \delta_{\mu 2} 
\right\}\; .
\ee
Here, m is the fermion mass, s is the CM energy square. From (\ref{e17}) 
we have
\be
\label{e18}
A_\mu (p, \lambda; p', \lambda') = 2 m\; \delta_{\mu 3}\;\;\; {\rm for}
\;\; \lambda = \lambda'\; , 
\ee
\be
A_\mu (p, \lambda; p', \lambda') = - \lambda \sqrt{s - 4 m^2}\; 
\delta_{\mu 1} + {\rm I} \sqrt{s}\; \delta_{\mu 2} \;\;\;
{\rm for}\;\; \lambda = - \lambda'\; ,
\ee
which are exactly the same as given in eq. (23) of Ref. \cite{vewu} 
after taking a specific momentum direction there.

In summary, we provide a new approach for evaluating Feynman diagrams 
symbolically, which is a straightforward and efficient promotion of 
the methods developed in Refs.\cite{ks,bama}. We have presented 
examples in displaying the simpleness of this approach, and the new 
method is applicable to both massive and massless fermion amplitude 
calculations. Especially, since there is no restriction on the working 
frame, or in other words the fermion momentum directions, by this 
approach, it is quite convenient to encode the helicity amplitudes 
into a symbolic manipulating program for computer. Another
advantage of the method is that the expression (\ref{e8}) is Lorentz
index covariant, although one needs to specify the auxiliary vectors 
in the end for one's convenience, which makes the intermediate steps 
of calculation not so tedious. In the end, we would like to mention
that most phenomenological applications of massive fermion amplitudes 
can be improved by also including the decay of the fermions. It is 
often possible to factorize the problem into production and decay 
amplitudes, and the new technique could be used for both in principle. 
However, to do this much care should be paid to the helicity state 
conventions.

\vspace{10mm}

\centerline{\large \bf Acknowledgments} 

\noindent
The author would like to thank the ICTP for hospitality, 
while part of this work was carried out.


\begin{thebibliography}{99}
\bibitem{r1} J.L. Powell, Phys.\ Rev.\ {\bf 75},\ 32 (1949).

\bibitem{r2} M. Jacob and G. Wick, Ann.\ Phys.\ (N.Y.)\ {\bf 7},\ 404
(1959).

\bibitem{r3} J. Bjorken and M. Chen, Phys.\ Rev.\ {\bf 154},\ 1335 
(1966).

\bibitem{r4} P. De Causmaecker, R. Gastmans, W. Troosts and T.T. Wu,
Phys. Lett. B{\bf 105}, 215 (1981); Nucl. Phys. B{\bf 206}, 53 (1982).

\bibitem{r5} J.F. Gunion and Z. Kunszt, Phys. Lett. 
B{\bf 161} 333 (1985).

\bibitem{r6} G. Farar and F. Neri, Phys. Lett. B{\bf 130} 109 (1983).

\bibitem{care}
M. Caffo and E. Remiddi, \HPA{55} (1982) 339.
 
\bibitem{pass}
G. Passarino, \PR{D28} (1983) 2867, \NP{B237} (1984) 249.

\bibitem{hz} K. Hagiwara and D. Zeppenfeld,
\NP{B274} (1986) 1.
 
\bibitem{ks} F.A. Berends, P.H. Daverveldt and R. Kleiss
\NP{B253} (1985) 441, R. Kleiss and W.J. Stirling,
\NP{B262} (1985) 235.

\bibitem{bama} A. Ballestrero and E. Maina
\PL{B350}, 225 (1995).
 
\bibitem{xuzc} Z. Xu, Da-Hua Zhang and L. Chang
\NP{B291} (1987) 392.
 
\bibitem{vewu} R. Vega and J. Wudka,
\PR{D53}, 5286 (1996).

\bibitem{mana} C. Mana and M. Martinez,
\NP{B287} (1987) 601.
 
\bibitem{bomi} C. Bouchiat and L. Michel,
\NP{5} (1958) 416.

\bibitem{peskin} M.E. Peskin and D.V. Schroeder, {\it An Introduction
to Quantum Field Theory} (Addison-Wesley, USA, 1995). 

\end{thebibliography}
\end{document}